\documentstyle[prl,aps,multicol]{revtex}
\renewcommand{\narrowtext}{\begin{multicols}{2}
\global\columnwidth20.5pc}
\renewcommand{\widetext}{\end{multicols} \global\columnwidth42.5pc}
\def\Lrule{\vspace*{-0.2in}\noindent\vrule width3.5in height.2pt
  depth.2pt \vrule depth0em height1em}
\def\Rrule{\vspace{-0.1in}\hfill\vrule depth1em height0pt \vrule
  width3.5in height.2pt depth.2pt\vspace*{-0.1in}}
\def\bml{\begin{mathletters}}
\def\eml{\end{mathletters}}
\def\beq{\begin{equation}}
\def\eeq{\end{equation}}
\def\bea{\begin{eqnarray}}
\def\eea{\end{eqnarray}}
\def\bey{\begin{eqnarray*}}
\def\eey{\end{eqnarray*}}
\def\ba{\begin{array}}
\def\ea{\end{array}}
\def\to{\rightarrow}

\begin{document}
\preprint{SUNY-NTG-00/55, TIT-HEP-456}
\draft
\title{
Chiral Lagrangians from lattice gauge theories
in the strong coupling limit}
\author{Taro Nagao}
\address{
Department of Physics, Graduate School of Science,
Osaka University, Toyonaka, Osaka 560-0043, Japan}
\author{Shinsuke M. Nishigaki${}^*$}
\address{
Department of Physics \& Astronomy,
State University of New  York, Stony Brook, NY 11794-3800, USA
\ {\rm and}\\
Department of Physics, Tokyo Institute of Technology,
O-okayama, Meguro, Tokyo 152-8551, Japan}
\date{December 29, 2000}
\maketitle
\begin{abstract}
We derive nonlinear $\sigma$ models (chiral Lagrangians)
over symmetric spaces
${\rm U}(n)$, ${\rm U}(2n)/{\rm Sp}(2n)$, and ${\rm U}(2n)/{\rm O}(2n)$
from ${\rm U}(N)$, ${\rm O}(N)$, and ${\rm Sp}(2N)$ lattice gauge 
theories coupled to $n$ flavors of staggered fermions, 
in the large-$N$ and $g^2N$ limit.
To this end, we employ Zirnbauer's color-flavor transformation. 
We prove the spatial homogeneity of the vacuum configurations
of mesons by explicitly solving the large-$N$ saddle point equations,
and thus establish the above patterns of spontaneous 
chiral symmetry breaking
without any assumptions.
\end{abstract}
\pacs{PACS number(s): 11.15.Ha, 11.15.Me, 11.15.Pg, 11.30.Rd}
\setcounter{footnote}{0}
\narrowtext
{\it Introduction.}
The concept of spontaneous symmetry breaking 
introduces an essential parallelism between 
the spectral statistics of disordered condensate 
\cite{Efe} and the low-energy dynamics of strongly 
coupled gauge field theories \cite{Wei}, involving 
a juxtaposition of the lightest and 
weakly coupled Goldstone particles in these theories: 
diffusion modes associated with spontaneous breakdown of the 
symmetry between advanced and retarded Green's functions, and 
pions associated with
spontaneous breakdown of the chiral symmetry. 
However there is a considerable difference
at the practical level.
An ensemble averaging can be analytically performed
for disordered Hamiltonians,
so that the Goldstone manifold and the nonlinear $\sigma$ model
over it describing the low-frequency regime of the theory 
manifest themselves without ambiguity. On the other hand, 
gauge theories with propagating gluons do not admit
such an analytic computation.
Accordingly, in order to identify a low-energy effective theory
one needs to appeal to nonperturbative theorems, namely 
Vafa-Witten's theorem \cite{VW} that forbids spontaneous
breaking of the vector symmetry in a vectorial theory,
and 't Hooft's anomaly matching \cite{tHo}
between fundamental and effective theories. 
Relying on these theorems, non-standard patterns 
of chiral symmetry breaking have been studied
\cite{Pes} in the context of technicolor models.
\linebreak
\indent
The symmetry-based approach puts a strong restriction on
operators that are allowed in the effective chiral Lagrangian, 
but the setback is that their coupling constants are left as
free parameters that can by no means be related to 
those of the fundamental theory.
To overcome this difficulty, attempts \cite{KS} were 
made to derive an effective theory directly 
from a microscopic model under an extreme condition,
namely a ${\rm U}(N)$ lattice gauge theory
in the strong coupling limit, combined with the large-$N$ limit. 
This serves as a basis on which 
a strong coupling expansion is performed and 
the plaquette action can be taken into account
perturbatively in $(g^2N)^{-1}$ \cite{Jol}.

Recently Altland and Simons \cite{AS} have renewed interest
in this subject.
Instead of computing a one-link integral via the 
Br\'{e}zin-Gross formula \cite{BG} valid only in the large-$N$ limit,
they have adopted an alternative method,
Zirnbauer's color-flavor transformation \cite{Zir2,Zir3},
which has a clear advantage of being valid at finite $N$. 
By converting the integrations over strongly coupled gluonic variables
into supposedly weakly coupled mesonic variables, they have
provided a transparent derivation of  
the low-energy effective action corresponding to 
fermions in the complex representation. 
However, their and all the previous works have relied on
a crucial and far-from-evident assumption that the mesonic 
fields in the vacuum are homogeneous.

The purpose of this paper is twofold:
by applying the color-flavor transformation to strongly coupled
lattice gauge theories with three classical gauge groups
and with staggered fermions, we shall
i) demonstrate
the spatial homogeneity of the mesonic fields in the vacuum,
thereby establishing the chiral symmetry breaking rigorously,
ii) extend the method of Altland and Simons
valid for quarks in the complex representation of the gauge group,
to other two classes of representations, so as to derive
non-standard type of chiral Lagrangians \cite{KST}.

{\it Lattice Gauge Theory with Staggered Fermions.}
We consider a lattice gauge theory coupled to
staggered fermions on an even($=d$) dimensional lattice 
${\bf Z}^d \ni x$,
which is split into
odd and even sublattices labeled by
$\circ$ and $\times$, respectively.
The set of $2d$ odd (even) sites adjacent to 
an even (odd) site $x$
will be denoted as $\circ_x$ ($\times_x$).
We write the partition function
with an emphasis on the bipartite nature
(the lattice spacing is set to unity):
\widetext
\Lrule
\bea
&&
{\cal Z}(g;m)=
\int_{G}
\prod_{x \in \times , y \in \circ_x}
\!\!\!\!\!\!
dU_{x{}y}
\int
\!\!
\prod_{x\in \times, \circ}
\!\!
d\bar{\psi}_x \, d{\psi}_x\,
\exp\left[
S[U]
+\!\!\!\!
\sum_{x\in \times , y \in \circ}
\!\!
\frac{\eta_{xy}}{2}
\left(
 \bar{\psi}^{i,a}_{x} U^{ij}_{x{}y} {\psi}^{j,a}_y
-\bar{\psi}^{j,a}_{y} U_{x{}y}^{\dagger ji}
{\psi}^{i,a}_{x}
\right)
-\!\!\!\!
\sum_{x\in \times, \circ}   \bar{\psi}^{i,a}_x m^{ab} {\psi}^{i,b}_x
\right].
\label{ZQCD}
\eea
\narrowtext
\noindent
Here the link variable
$U^{ij}_{x{}y}$ takes its value in the gauge group
$G={\rm U}(N), {\rm O}(N)$, or ${\rm Sp}(2N)$.
The Haar measure of $G$ is denoted by $dU$,
and $S[U]=g^{-2}\sum_{\langle xyzw \rangle}
 {\rm tr}\,U_{xy}U^\dagger_{zy}U_{zw}U^\dagger_{xw}+{\rm c.c.}$ 
is the plaquette gauge action.
The site variables ${\psi}^{i,a}_x$ and ${\bar{\psi}}^{i,a}_x$
are independent Grassmannian numbers.
The color indices $i,j,\ldots$ run from 1 
to $N$ for $G={\rm U}(N), {\rm O}(N)$ or to $2N$ for $G={\rm Sp}(2N)$,
and
the flavor indices $a,b,\ldots$ run from 1 to $n$.
The phase is defined by
$\eta_{x\,x\pm\hat{\mu}}=(-1)^{x^1+\cdots+x^{\mu-1}}$, and
the quark mass matrix is by $m={\rm diag}(m_1,\ldots,m_n)$.

As the fundamental representation of 
${\rm O}(N)$ (${\rm Sp}(2N)$) is real (pseudoreal),
i.e. its symmetric (antisymmetric) product contains an invariant,
the chiral symmetry group is extended to
${\rm U}(2n)$ for these cases \cite{Pes}.
This can be made apparent by introducing $2n$-flavored site variables
\bea
&&
\Psi^{i,A}_y = 
[
\psi^{i,a}_y \ \bar{\psi}_y^{i,a}
]
\ \ \mbox{or}\ \ 
[
\psi^{i,a}_y \ {E}^{ij}\bar{\psi}_y^{j,a}
]
\ (y\in\circ) ,
\\  
&&
\bar{\Psi}^{i,A}_x = 
[
\bar{\psi}^{i,a}_x \ {\psi}_x^{i,a}
]
\ \ \mbox{or}\ \ 
[
\bar{\psi}^{i,a}_x \ E^{ij} {\psi}_x^{j,a} 
]
\ (x\in\times),
\nonumber
\eea
$(A,B=1,\ldots,2n, E=i\sigma_2 \otimes \openone_N)$
for $G={\rm O}(N)$ or ${\rm Sp}(2N)$, resp.,
and writing the fermionic action as
\widetext
\Lrule
\bml
\bea
&&\sum_{x\in \times, y \in \circ_x}
\frac{\eta_{xy}}{2} 
\bar{\Psi}^{i,A}_{x} O^{ij}_{x{}y} {\Psi}^{j,A}_y 
-\frac12 \sum_{y\in \circ} \Psi_{y}^{A,i} {m}^{AB} \Psi_{y}^{B,i} 
+\frac12 \sum_{x\in \times} \bar{\Psi}_{x}^{A,i} {m}^{AB} 
\bar{\Psi}_{x}^{B,i} 
\ \ 
\bigl(O_{xy}\in{\rm O}(N)\bigr),\\ 
&&\sum_{x\in \times, y \in \circ_x}
\frac{\eta_{xy}}{2} 
\bar{\Psi}^{i,A}_{x} S^{ij}_{x{}y} {\Psi}^{j,A}_y 
-\frac12 \sum_{y\in \circ} \Psi_{y}^{A,i} {E}^{ij} {m}^{AB} 
\Psi_{y}^{B,i} 
+\frac12 \sum_{x\in \times} \bar{\Psi}_{x}^{A,i} {E}^{ij}{m}^{AB} 
\bar{\Psi}_{x}^{B,i} 
\ \ 
\bigl(S_{xy}\in{\rm Sp}(2N)\bigr).
\eea
\label{ext}
\eml
\Rrule
\narrowtext
\noindent
Extended mass matrices ${m}^{AB}$ 
in the above are defined 
\nolinebreak[3] by
$
{m}=
(i\sigma_2\  \mbox{or}\ \sigma_1)
\otimes 
{\rm diag}(m_1,\ldots,m_n)
$
for  $G={\rm O}(N)$ or ${\rm Sp}(2N)$, respectively.
The protected vector subgroup of 
this extended chiral group ${\rm U}(2n)$
is read off from Eq.(\ref{ext}) to be
${\rm Sp}(2n)$ for $G={\rm O}(N)$, or
${\rm O}(2n)$ for $G={\rm Sp}(2N)$.

{\it Color-Flavor Transformation.}
Henceforth we concentrate on the strong coupling limit 
$g^2 N \to \infty$.
The theory is thus referred to as a non-Abelian random flux model.
In this case, 
$\eta_{x{}y}$ can be absorbed into the redefinition of
the link variables.
We can reexpress the partition function of this model
in terms of an integration over flavor-singlet link variables
into the one over color-singlet link variables,
by Zirnbauer's color-flavor transformation.
That for the ${\rm U}(N)$ link variable coupled to fermions reads
\cite{Zir2,Zir3}, up to an irrelevant numerical factor,
\widetext
\Lrule
\bml
\beq
\int_{{\rm U}(N)}\!\!\!\!\!
dU\,
\exp\left(
\bar{\psi}_1^{i,a} U^{ij} \psi_0^{j,a}-
\bar{\psi}_0^{j,a} U^{\dagger ji} \psi_1^{i,a}
\right)
=
\int
\frac{\prod\nolimits_{1\leq a,b\leq n} 
dZ^{ab}d\bar{Z}^{ab}}{\det{}^{N+2n}(1+Z^\dagger Z)}
\exp\left(
\bar{\psi}^{i,a}_{1} Z^{ab} \psi^{i,b}_1+
\bar{\psi}^{i,a}_{0} Z^{\dagger ab} \psi^{i,b}_0
\right) .
\eeq
The integration on the RHS is over complex $n\times n$ 
matrices, ${\cal N}={\rm GL}(n,{\bf C})$.
Although the case with the ${\rm O}(N)$ link variable 
coupled to fermions was not explicitly provided
(see Ref.\cite{Zir3} for its bosonic counterpart), 
one can show that
\beq
\int_{{\rm O}(N)}\!\!\!\!\!
dO\,
\exp\left(
\bar{\Psi}^{i,A} O^{ij} \Psi^{j,A}
\right)
=
\int_{
Z=-Z^T
}
\frac{\prod\nolimits_{1\leq a<b\leq 2n} 
dZ^{ab}d\bar{Z}^{ab}}{\det{}^{N/2+2n-1}
(1+Z^\dagger Z)}
\exp\left(
\bar{\Psi}^{i,A} Z^{AB} \bar{\Psi}^{i,B}
-{\Psi}^{i,A} Z^{\dagger AB} \Psi^{i,B}
\right).
\eeq
The integration on the RHS is over the set of 
complex $2n\times 2n$ antisymmetric
matrices, which is diffeomorphic to
${\cal N}={\rm GL}(2n,{\bf C})/{\rm Sp}(2n,{\bf C})$.
The case with the ${\rm Sp}(2N)$ link variable 
coupled to fermions reads \cite{Zir2}
\beq
\int_{{\rm Sp}(2N)}\!\!\!\!\!
dS\,
\exp\left(
\bar{\Psi}^{i,A} S^{ij} \Psi^{j,A}
\right)
=
\int_{
Z=Z^T}
\frac{\prod\nolimits_{1\leq a\leq b\leq 2n} 
dZ^{ab}d\bar{Z}^{ab}}{\det{}^{N+2n+1}
(1+Z^\dagger Z)}
\exp\left(
\frac12 \bar{\Psi}^{i,A} E^{ij} Z^{AB} \bar{\Psi}^{j,B}-
\frac12 {\Psi}^{i,A} E^{ij} Z^{\dagger AB} \Psi^{j,B}
\right).
\eeq
\eml
The integration on the RHS is over the set of 
complex $2n\times 2n$ symmetric
matrices, 
which is diffeomorphic to
${\cal N}={\rm GL}(2n,{\bf C})/{\rm O}(2n,{\bf C})$.
Applying these transformations to each link
and then integrating over fermions,
we obtain 
\beq
{\cal Z}
(m)=
\int_{{\cal N}}
\prod_{x\in \times,\ y\in \circ_x} 
\frac{dZ_{xy}\,d{\bar{Z}}_{xy}}{\det^{N+2n+\epsilon} 
(1+Z^\dagger_{xy}Z_{xy})}
\prod_{x\in \times}\det\nolimits^N\left(\sum_{y\in \circ_x}Z_{xy} -2m 
\right)
\prod_{y\in \circ}
\det\nolimits^N\left(\sum_{x\in \times_y}Z^\dagger_{xy} -2m \right).
\eeq
Here $\epsilon=0, -1, 1$ for $G={\rm U}(N)$, ${\rm O}(N)$, ${\rm Sp}(2N)$,
respectively, and
we have redefined $N/2\to N$ and $m\to 2m$ for the 
${\rm O}(N)$ case.
Pfaffians from fermionic integrations 
${\rm Pf}(X)=\int d\Psi \exp(\Psi X \Psi)$
arising in the ${\rm O}(N)$ and ${\rm Sp}(2N)$ cases
are replaced by $\det{}^{1/2}(X)$.
In spite of residing on a link,
$Z_{xy}$ variables resemble the mass matrix
that couples to a fermions bilinear on an identical site,
thereby indicating the spontaneous chiral symmetry breaking.

{\it Homogeneity of the Saddle Point.}
So far the manipulation applied to our non-Abelian random flux models 
was exact. 
We finally make an approximation, by taking the $N\to\infty$ limit.
The saddle point equations read
\narrowtext
\beq
Z_{xy}^{-1}=
\sum_{x'\in \times_y\atop
x'\neq x} Z^\dagger_{x'y}-2m ,\ 
(Z^{\dagger}_{xy})^{-1}=
\sum_{y'\in \circ_x\atop  
y'\neq y } Z_{xy'} -2m
\label{saddle}
\eeq
($x\in\times,\ y\in\circ_x$) for $G={\rm U}(N)$. 
Those for $G={\rm O}(N)$ or ${\rm Sp}(2N)$ are
identical as the above,
with $Z_{xy}$ constrained to be antisymmetric or symmetric.

Now we shall demonstrate that the unique solution to Eq.(\ref{saddle}) 
at $m=0$ is given by
\bea
&&Z_{xy} =(2d-1)^{-1/2}U, \ U\in {\rm U}(n) 
\ \mbox{for}\  {}^\forall x \in \times, {}^\forall y\in \circ_x,
\label{hom}
\eea
i.e.\ the vacuum configuration of the mesonic variable is homogeneous. 
In other two cases, we simply need to restrict
$U$ to be an antisymmetric ($G={\rm O}(N)$) or
a symmetric ($G={\rm Sp}(2N)$) unitary matrix, which in turn can be 
parametrized by another unitary matrix $V\in {\rm U}(2n)$ 
by the canonical projection $U=V J V^T$ 
$(J=i\sigma_2 \otimes \openone_n)$
or $U=V V^T$, respectively.

The set of saddle point equations involving an even site $x$ read,
after using abbreviations
$Z_p\equiv Z_{x y_p}$
for $y_1,\ldots,y_{2d}\in\circ_x$,
\begin{equation}
(Z_p^\dagger)^{-1}= \sum_{q=1 
\, 
({q} \neq p)}^{2d} Z_q, \ \ 
p=1,{}\ldots,{{2d}}(\geq 4).
\label{1.1}
\end{equation}
Here
$Z_1,\ldots,Z_{2d} \in {\rm GL}(n,{\bf C})$. 
Eq.(\ref{1.1}) clearly leads to
\begin{equation}
(Z_p^{\dagger})^{-1} + Z_p = 
(Z_q^{\dagger})^{-1} + Z_q, \ \ 
\forall p, \forall q=1,{}\ldots,2d.  
\label{1.2}
\end{equation} 
We make singular value decomposition of the matrices
\beq
Z_p = U_p S_p V_p, \ \ S_p=
{\rm diag}( s^{(p)}_1 , \ldots, s^{(p)}_n ),
\eeq
where $U_p, V_p \in {\rm U}(n)$
and $s^{(p)}_1,\ldots,s^{(p)}_n > 0$,
and define
$
\sigma^{(p)}_a = 1/{s^{(p)}_a} + s^{(p)}_a
$. 
Without loss of generality we choose 
$(U_p, V_p)$ so that 
$
\sigma^{(p)}_1 \geq \cdots \geq \sigma^{(p)}_n
$.   
{}From Eq.(\ref{1.2}), we have 
\bea
U_p(S_p^{-1} + S_p)V_p &=& U_q(S_q^{-1} + S_q)V_q 
\nonumber
\\
\Rightarrow
U_q^{\dagger} U_p (S_p^{-1} + S_p) V_p V_q^{\dagger} &=& 
V_q V_p^{\dagger} (S_p^{-1} + S_p) U_p^{\dagger} U_q 
\nonumber\\
&=& S_q^{-1} + S_q
\label{1.7}\\
\Rightarrow 
U_q^{\dagger} U_p (S_p^{-1} + S_p)^2  U_p^{\dagger} U_q
 &=& (S_q^{-1} + S_q)^2.
\label{1.8}
\eea
{}From the uniqueness of the eigenvalues, we find 
that the diagonal matrix $S_p^{-1} + S_p$ is independent of $p$, 
and thus is simply denoted as 
$
S_p^{-1} + S_p 
 = {\rm diag}\left( 
\sigma_1 , \ldots, \sigma_n  \right).
$
Then Eq.(\ref{1.8}) in a componentwise notation reads
\begin{equation} 
(U_q^{\dagger} U_p)^{ab} ( \sigma_b^2 - \sigma_a^2 ) 
= 0, \ \ \ a,b= 1,{}\ldots,n.
\end{equation} 
As $\sigma_b + \sigma_a$ is nonzero, we find
\bea
&&(U_q^{\dagger} U_p)^{ab} (\sigma_b - \sigma_a) = 0, \ \ \ 
a,b = 1,{}\ldots,n
\label{1.11}\\
\Leftrightarrow&&
U_q^{\dagger} U_p (S_p^{-1} + S_p)  = (S_p^{-1} + S_p) U_q^{\dagger} 
U_p.
\label{1.12}
\eea 
Using Eq.(\ref{1.12}) in Eq.(\ref{1.7}), we obtain
\begin{equation}
U_q^{\dagger} U_p = V_q V_p^{\dagger}
\equiv W_{qp} .
\label{1.13}
\end{equation}
Note that $W_{qp}^{\dagger} = W_{pq}$. 
In terms of $W_{qp}$, Eq.(\ref{1.1}) reads
\begin{equation}
S_p^{-1}= \sum_{{q}=1 
\,(
{q} \neq p)}^{2d} 
W_{qp}^\dagger S_{{q}} W_{{q} p} , \ \ \ p=1,{}\ldots,{2d}.
\label{1.14}
\end{equation}
Let us now suppose
$
\sigma_1 =  \cdots = \sigma_r > \sigma_{r+1} \geq \cdots 
\geq \sigma_n
$
and consider possible two cases separately.

In the case $r=n$,
we have
\begin{equation}
S_p^{-1} + S_p = 
\sigma \openone_n,
\ \ p=1,\ldots,2d.
\label{1.18}
\end{equation}
We rewrite the saddle point equations (\ref{1.1}) as 
\begin{equation}
\sum_{{r}=1
\, ({r} \neq p,q)}^{2d} 
W_{{r} p}^{\dagger} S_{{r}} W_{{r} p} 
= S_p^{-1} - W_{qp}^{\dagger} S_q W_{qp},  
\label{1.19}
\end{equation}
where $p,q = 1,\ldots,{2d}$ and $p \neq q$. 
Since 
$W_{{r} p}^{\dagger} S_{{r}} W_{{r} p}$ are positive-definite, 
the LHS of Eq.(\ref{1.19}) is positive-definite. Therefore $S_p^{-1}$ 
and $W_{qp}^{\dagger} S_q W_{qp}$ are ordered hermitian matrices 
\cite{Meh} so that the largest 
and smallest diagonal elements (eigenvalues) of $S_p^{-1}$ are larger 
than 
the largest and smallest diagonal elements of $S_q$, respectively. 
Suppose the 
largest diagonal element of $S_p^{-1}$ is $s$. From Eq.(\ref{1.18}),  
the diagonal elements of $S_p$ and $S_q$ are $s$ or $1/s$. The 
largest diagonal element of $S_q$ is $1/s$ and $s>1$, since it must be 
smaller than $s$. 
Therefore all the diagonal elements of $S_q$ are $1/s$. 
Then the smallest diagonal element of $S_p^{-1}$ has to be $s$, 
because it must be larger than the smallest diagonal element of $S_q$. 
Consequently all the diagonal elements of $S_p^{-1}$ are $s$ and 
we conclude
\begin{equation}
Z_p = U_p S_p V_p = U_p V_p S_p = U_q V_q S_q = U_q S_q V_q = Z_q.
\end{equation}

In the case $r<n$, we obtain from Eqs.(\ref{1.11}) and (\ref{1.13}), 
\beq
(W_{qp})^{ab} = 0
\ \ \mbox{for}\ 
\left\{
\ba{l}
a = 1,\ldots,r, \ \ b=r+1,\ldots,n \\
a=r+1,\ldots,n, \  \ b= 1,{}\ldots,r
\ea \right.
\eeq
for an arbitrary pair of indices $p$ and $q$.
It means that the saddle point equation (\ref{1.14}) 
is decomposed into those with smaller ranks $r$ and $n-r$.
Thus we can inductively show $Z_1 = \cdots = Z_{2d}$,
using the argument given in the previous case
for each irreducible component.

Substituting $Z_1 = \cdots = Z_{2d}$ into Eq.(\ref{1.1}), we obtain
\beq
Z_p=(2d-1)^{-1/2} U,
\ \ U\in {\rm U}(n),
\ \ \forall p=1,\ldots,2d.
\eeq
As this proof applies for each of even sites, or odd sites 
with $Z_{xy}$ substituted by $Z^\dagger_{xy}$,
we finally establish Eq.(\ref{hom}).

{\it Chiral Lagrangian.}
Next we take into account 
a small deviation from the chiral limit, as well as
fluctuations of the mesonic field around the vacuum
configuration that has been proven in the last section
to be homogeneous.

The terms to be collected are
of lowest nontrivial orders in masses
and momenta, which are $O(m^1)$ and $O(\partial^2)$, respectively.
In the case of $G={\rm U}(N)$ \cite{AS}, we split the
mesonic variable into massive and Goldstone modes as
\bml
\beq
Z_{xy}=(2d-1)^{-1/2} P_{xy} \,U_{xy},
\eeq
where $P_{xy}\in {\rm GL}(n, {\bf C})/{\rm U}(n)$ and 
$U_{xy}\in {\rm U}(n)$.
In the case of $G={\rm O}(N)$ or ${\rm Sp}(2N)$, the mesonic variable 
is constrained to be antisymmetric or symmetric, respectively, 
so we must employ parameterizations
\bea
&&Z_{xy}=(2d-1)^{-1/2} P_{xy} V_{xy} J V_{xy}^T P^T_{xy}
\ \ 
\bigl( G={\rm O}(N) \bigr),\!
\\
&&
Z_{xy}=(2d-1)^{-1/2} P_{xy} V_{xy} V_{xy}^T P^T_{xy}
\ \ 
\bigl( G={\rm Sp}(2N) \bigr),\!
\eea
\eml
where
$V_{xy}$ is the Goldstone mode
over ${\cal M}={\rm U}(2n)/$ ${\rm Sp}(2n)$ or 
${\rm U}(2n)/{\rm O}(2n)$, resp., and $P_{xy}$ 
is the massive mode over ${\cal N}/{\cal M}$.
Then we expand the Goldstone mode
$
U_{xy}=\exp({\partial_{\vec{xy}}}) U_{x}
$
up to quadratic order in the directional derivatives
$\partial_{\vec{xy}}$,
and Gaussian-integrate over the massive mode
$
P_{xy}=\exp({X_{xy}})
$
by retaining up to quadratic order in $X_{xy}$.
We reinstate the lattice spacing $a$
and employ dimensionful continuum notations
\[
m\to a M ,\ \ 
\sum_{y\in \times_x} \partial^2_{\vec{xy}} \to \frac12 a^2 \nabla^2,\ \ 
\sum_{x\in \times}\to \frac12 a^{-d} \int d^d x .
\]
Then the resulting effective partition function reads
\bea
&&{\cal Z}
(M)=\int_{{\cal M}}
\!\!{\cal D}V\,
\exp\left[
-\int d^d x
\left(
 \frac{N}{8d} a^{2-d}{\rm tr}\,\nabla U^\dagger \nabla U
\right.\right.
\label{chL} \!\\
&&
\ \ \ \left.\left.
- \frac{N \sqrt{2d-1}}{d}  a^{1-d} {\rm Re\,tr}\,M U
+C a^{2-d} ({\rm tr}\,U^\dagger \nabla U)^2
\right)
\right],
\nonumber
\eea
where 
$U=V, VJV^T, VV^T$ 
for $G={\rm U}(N)$, ${\rm O}(N)$, ${\rm Sp}(2N)$,
respectively, and ${\cal D}V=\prod_{x\in \times}dV_x$.
The coefficient $C$ of the singlet ($\eta'$) kinetic term
is a constant of $O(1)$ 
depending on $d$ and $G$ \cite{AS},
but we suppress its explicit form
as $\eta'$ should decouple from the Goldstone sector if
the ${\rm U}(1)_A$ anomaly were properly taken into account.
Consequently 
the pion decay constant squared
and the chiral condensate
(the coefficients of the first and the second terms)
are derived from the first principle.

{\it Conclusion.}
We have rigorously derived three classes of 
low-energy chiral Lagrangians from microscopic theories: 
non-Abelian random flux models,
or lattice gauge theories with
staggered fermions in the strong coupling limit.
These nonlinear $\sigma$ models have
symmetric spaces ${\cal M}={\rm U}(n), {\rm U}(2n)/{\rm Sp}(2n), 
{\rm U}(2n)/{\rm O}(2n)$ as their target manifolds, depending on the 
complexity, reality, and pseudoreality of the defining representations 
of the gauge groups
$G={\rm U}(N), {\rm O}(N), {\rm Sp}(2N)$, respectively.
These results are anticipated from Vafa-Witten argument \cite{KST}
or equivalently from the anti-unitary symmetries
of the staggered Dirac operators in Eq.(\ref{ZQCD}),
whose matrix  elements are trivially complex, real, and quaternion 
real for these three gauge groups.
For the last two cases of the gauge groups,
the anti-unitary symmetries of lattice Dirac operators
are interchanged from those of the continuum Dirac operators,
due to the absence of the charge conjugation matrix
acting on spinor indices, that squares to $-1$ \cite{HVlat}.
The puzzle of if and how the crossover between different classes of
effective theories could occur, originally posed for staggered fermions in 
the adjoint of ${\rm SU}(2)$ ($=$ fundamental of ${\rm O}(3)$) and in
the fundamental of ${\rm SU}(2)$ ($\approx$ fundamental of 
${\rm Sp}(2)$), remains unsolved.
Although the Wess-Zumino term in $d=2$ dimensions
is of the same order as the kinetic term,
our models involving staggered fermions do not yield it.

Verbaarschot \cite{Ver} has conjectured 
that the three patterns of chiral symmetry breaking induce
the spectral fluctuation of the Dirac operators
to obey three universality classes of chiral random matrices \cite{ShV}.
If our parameters are restricted to the `ergodic' domain
where the size of the lattice is
much smaller than the Compton length of pseudo-Goldstone bosons,
the chiral Lagrangians are dominated by their zero modes,
i.e. become finite-dimensional integrals involving only the mass terms
in Eq.(\ref{chL}) \cite{LS}.
As these three `finite-volume' partition functions are known 
\cite{ShV,HV} to be equivalent to 
chiral random matrix ensembles at $\beta=2,1,4$
in the large matrix-dimension limit,
we have also proven his $\mbox{conjecture}$.

\widetext
\end{document}